\documentclass[useAMS,usenatbib,usegraphicx]{mn2e}
\title[Discovery of a magnetic field in the O9 sub-giant star HD~57682 by the MiMeS Collaboration]{Discovery of a magnetic field in the O9 sub-giant star HD~57682 by the MiMeS Collaboration\thanks{Based on observations obtained at the Canada-France-Hawaii Telescope (CFHT) which is operated by the National Research Council of Canada, the Institut National des Sciences de l'Univers of the Centre National de la Recherche Scientifique of France,  and the University of Hawaii.}}
\author[Grunhut et al.]{J.H. Grunhut$^{1,2}$\thanks{E-mail: Jason.Grunhut@rmc.ca}, G.A. Wade$^2$, W.L.F. Marcolino$^3$, V. Petit$^{4}$, H.F. Henrichs$^5$,
\newauthor
D.H. Cohen$^6$, E. Alecian$^7$, D. Bohlender$^8$, J.-C. Bouret$^3$, O. Kochukhov$^9$, 
\newauthor
C. Neiner$^{10}$, N. St-Louis$^{11}$, R.H.D. Townsend$^{12}$, and the MiMeS Collaboration\\
$^1$Department of Physics, Engineering Physics \& Astronomy, Queen's University, Kingston, Ontario, Canada, K7L 3N6\\
$^2$Department of Physics, Royal Military College of Canada, P.O. Box 17000, Station Forces, Kingston, Ontario, Canada, K7K 7B4\\
$^3$LAM-UMR 6110, CNRS \& Univ. de Provence, 38 rue Fr\'{e}d\'{e}ric Joliot-Curie, F-13388 Marseille cedex 13, France\\
$^4$D{\'e}partement de physique, g{\' e}nie physique et optique, CRAQ, Universit{\' e} Laval, Qu{\' e}bec, Canada, G1K 7P4\\
$^5$Astronomical Institute ``Anton Pannekoek", University of Amsterdam, Science Park 904, 1098 XH Amsterdam, Netherlands\\
$^6$Department of Physics and Astronomy, Swarthmore College, 500 College Ave., Swarthmore, PA 19081, USA\\
$^7$LESIA, Observatoire de Paris, CNRS, UPMC, Universit\'e Paris Diderot, 5 place Jules Janssen, 92195 Meudon Cedex, France\\
$^8$National Research Council of Canada, Herzberg Institute of Astrophysics, 5071 W. Saanich Rd., Victoria, BC V9E 2E7, Canada\\
$^9$Department of Physics and Astronomy, Uppsala University, Box 515, 751 20 Uppsala, Sweden\\
$^{10}$GEPI, Observatoire de Paris, CNRS, Universit\'e Paris Diderot, 5 place Jules Janssen, 92195 Meudon Cedex, France\\
$^{11}$D{\' e}partement de Physique, Universit{\' e} de Montr{\' e}al, CP 6128, Succ. Centre-Ville, Montr{\' e}al, QC H3C 3J7, Canada\\
$^{12}$Department of Astronomy, University of Wisconsin-Madison, 5534 Sterling Hall, 475 N Charter Street, Madison, WI 53706, USA \\
}

\begin{document}
\date{\today}
\pagerange{\pageref{firstpage}--\pageref{lastpage}} \pubyear{2009}
\maketitle
\label{firstpage}
\begin{abstract}
We report the detection of a strong, organised magnetic field in the O9IV star HD~57682, using spectropolarimetric observations obtained with ESPaDOnS at the 3.6-m Canada-France-Hawaii Telescope within the context of the Magnetism in Massive Stars (MiMeS) Large Program. From the fitting of our spectra using NLTE model atmospheres we determined that HD~57682 is a $17^{+19}_{-9}$~M$_{\odot}$ star with a radius of $7.0^{+2.4}_{-1.8}$~R$_\odot$, and a relatively low mass-loss rate of $1.4^{+3.1}_{-0.95}\times10^{-9}$~M$_{\odot}$\,yr$^{-1}$. The photospheric absorption lines are narrow, and we use the Fourier transform technique to infer $v\sin i=15\pm3$~km\,s$^{-1}$. This $v\sin i$ implies a maximum rotational period of 31.5 d, a value qualitatively consistent with the observed variability of the optical absorption and emission lines, as well as the Stokes~$V$ profiles and longitudinal field. Using a Bayesian analysis of the velocity-resolved Stokes~$V$ profiles to infer the magnetic field characteristics, we tentatively derive a dipole field strength of $1680^{+134}_{-356}$~G. The derived field strength and wind characteristics imply a wind that is strongly confined by the magnetic field.
\end{abstract}


\section{Introduction}
The phenomenon of magnetism in hot, massive OB stars is not well studied. To date, repeated detections of circular polarization within line profiles have firmly established the presence of magnetic fields in three O-type stars ($\theta^1$~Ori~C, HD~191612, and $\zeta$~Ori~A; Donati et al. 2002, 2006; Bouret et al. 2008) and a handful of early B-type stars (e.g. Donati et al. 2001; Neiner et al. 2003; Alecian et al. 2008; Petit et al. 2008; Silvester et al. 2009). Indications of magnetic fields have also been reported in a number of other OB stars (e.g. by Hubrig et al. 2008; 2009). Such objects represent vital targets for the study of stellar magnetism. Their strong, radiatively-driven winds couple to magnetic fields, generating complex and dynamic magnetospheric structures (e.g. ud-Doula \& Owocki 2002), which modify mass loss, and may enhance the shedding of angular momentum via magnetic braking (e.g. ud-Doula et al. 2008). The presence of even a relatively weak magnetic field can profoundly influence the evolution of massive stars and their feedback effects, such as mechanical energy deposition in the ISM and supernova explosions (e.g. Ekstrom et al. 2008). For these reasons, the Magnetism in Massive Stars (MiMeS) Large Program (Wade et al. 2009; Grunhut et al. 2009) is exploiting the unique spectropolarimetric characteristics of ESPaDOnS at the Canada-France-Hawaii Telescope (CFHT) and NARVAL at the T{\' e}lescope Bernard Lyot (TBL) to obtain critical missing information about the poorly-studied magnetic properties of these important stars.

One of the MiMeS targets is HD~57682, an O9 sub-giant star. HD~57682 is reported to be a runaway star (Comeron et al. 1998) with no known companions (e.g. de Wit et al. 2005; Turner et al. 2008) and shows no photometric variability (e.g. Balona 1992). However, HD~57682 exhibits variability of UV lines characteristic of magnetic OB stars (e.g. Schnerr et al. 2008), and this is the primary reason it was included in the MiMeS target list.

\section{Observations}\label{obs_sec}
Eleven high-resolution ($R\sim65,000$), broadband (370-1050~nm) circular polarization (Stokes~$V$) spectra of HD~57682 were obtained with the ESPaDOnS spectropolarimeter, mounted on the 3.6-m CFHT, as part of the Survey Component of the MiMeS Large Program. The spectra were acquired and reduced in a manner essentially identical to that described by Silvester et al. (2009). The observations were obtained during two separate ESPaDOnS runs. The first two observations were obtained on successive nights in 2008 Dec. (high extinction hindered our first observation), and an additional 9 observations were obtained in 2009 May on five different nights. On nights when more than one observation was obtained, the un-normalised spectra were co-added prior to analysis, yielding 7 independent observations; A 5$^{\rm th}$-order polynomial (or lower) was used for normalisation of each individual order. A complete summary of the ESPaDOnS observations is given in Table~\ref{obs_tab}.  

\begin{figure}
\includegraphics[width=3.2in]{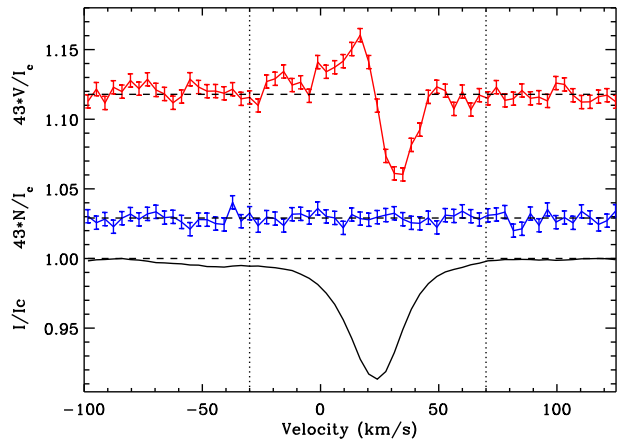}
\caption{LSD Stokes~$V$ (top), null $N$ (middle), and Stokes~$I$ profiles of HD~57682 from 2008 Dec. 6. The $V$ and $N$ profiles are expanded by the indicated factor and shifted upwards for display purposes. A clear Zeeman signature is detected in the Stokes~$V$ profile, while the null profile shows no signal. The integration limits used to measure the longitudinal field are indicated by the dotted lines.}
\label{lsd_examp}
\end{figure}

Least-Squares Deconvolution (LSD; Donati et al. 1997) was applied to all polarimetric observations in order to increase the signal-to-noise ratio (S/N) for this analysis. We adopted a mask in which all the Stark-broadened hydrogen and helium lines, and the metallic lines that are blended with the hydrogen lines were excluded. This resulted in approximately 165 relatively pure photospheric lines used for the creation of the mean, normalised Stokes $I$, $V$, and diagnostic null ($N$) profiles, yielding an increase in the Stokes~$V$ S/N by a factor of $\sim$7.5. All LSD profiles were computed on a spectral grid with a velocity bin of 3.6~km\,s$^{-1}$. The Stokes~$V$, diagnostic null, and the Stokes~$I$ profile of our 2008 Dec. 6 observation are shown in Fig.~\ref{lsd_examp} (a compilation of all Stokes $I$ and $V$ profiles is provided in Fig.~\ref{sv_fits}). A careful examination of the corresponding reduced spectrum clearly shows the presence of Stokes~$V$ signatures of similar morphology in numerous individual absorption lines of different chemical species. We also list the false alarm probabilities according to the criterion of Donati et al. (2002; 2006) in Table~\ref{obs_tab}, resulting in 5 definite detections, 1 marginal detection, and 1 non-detection.

The longitudinal magnetic field ($B_\ell$) was inferred from each LSD Stokes~$V$ profile in the manner described by Silvester et al. (2009). The longitudinal field measurements, which are reported in Table~\ref{obs_tab}, vary between -167~G and +282~G, with a typical uncertainty of 20-40~G. While the longitudinal field provides a useful statistical measure of the line-of-sight component of the field, we stress that we do not use it as the primary diagnostic of the presence of a magnetic field. This is because a large variety of magnetic configurations can produce a null longitudinal field. However, nearly all of these configurations will generate a detectable Stokes~$V$ signature in the velocity-resolved line profiles, which is what we observe.

In addition to the ESPaDOnS observations, archival IUE data were used to constrain the fundamental parameters of HD~57682 (see Sect.~\ref{fund_params}) and investigate its UV and optical line variability (see Sect.~\ref{temp_var_sec}). Three IUE spectra were obtained over a period of 5 years, one on 1978 Dec. 12 and the other two on 1983 Nov. 20. The first observation was obtained with the small aperture, while the latter were with the large aperture.

\begin{table}
\centering
\caption{Journal of ESPaDOnS observations listing the date, the heliocentric Julian date (2,454,000+), the number of sub-exposures and the exposure time per individual sub-exposure, and the peak signal-to-noise (S/N) ratio (per 1.8 km\,s$^{-1}$ velocity bin) in the Null spectrum in the $V$-band, for each night of observation. Columns 5-6 list the false alarm probability and the mean longitudinal field inferred from the LSD profiles.}
\begin{tabular}{@{}c@{\ \ \ }c@{\ \ \ }c@{\ \ \ }c@{\ \ \ }r@{\ }r@{}}
\hline
Date & HJD & $t_{\rm exp}$ & S/N & \multicolumn{1}{c}{FAP} & \multicolumn{1}{c}{$B_\ell \pm \sigma_B$} \\
\ & \ & (s) & \ & \ & \multicolumn{1}{c}{(G)} \\ 
\hline
2008/05/12 & 806.080 & $4\times500$ & 259 & $5\times10^{-4}$ & $282 \pm 114$      \\
2008/06/12 & 806.786 & $4\times500$ & 937 & $<10^{-8}$ & $266 \pm 30$     \\ 
2009/04/05 & 955.768 & $8\times600$ & 1502 & $3\times10^{-7}$ &$-46 \pm 21$    \\
2009/05/05 & 956.750 & $8\times540$ & 1401  & $1\times10^{-7}$  &$-35 \pm 22$  \\
2009/07/05 & 959.102 & $4\times540$ & 667  & $6\times10^{-2}$ &$-94 \pm 50$   \\
2009/08/05 & 959.749 & $8\times540$ & 1136 & $1\times10^{-8}$ &$-167 \pm 30$   \\ 
2009/09/05 & 960.748 & $8\times540$ & 915   & $6\times10^{-7}$ &$-109 \pm 37$  \\ 
\hline
\hline
\end{tabular}
\label{obs_tab}
\end{table}

\section{Stellar Parameters}\label{fund_params}
\begin{figure}
\centering
\includegraphics[width=3.3in]{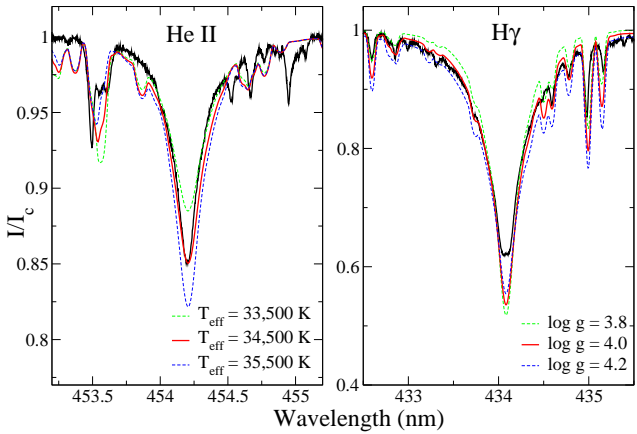}
\caption{Modelling of the He~{\sc ii} 454-nm line with atmospheric models corresponding to different $T_{\rm eff}$ values (left panel; $\log g=4.0$), and the H$\gamma$ line with models corresponding to different $\log g$ values (right panel; $T_{\rm eff}=34,500$~K) for the night of 2009 May 4. Note that only the wings of the H$\gamma$ profile are used; the core is likely contaminated by circumstellar material (see Sect.~\ref{disc_sec}).}
\label{parms_fig}
\end{figure}

Non-LTE, expanding atmosphere models calculated with the CMFGEN code were used in our analysis (see Hiller \& Miller 1998 for details). A summary of the fundamental parameters obtained is reported in Table~\ref{params}. The surface gravity ($\log g$) and effective temperature ($T_{\rm eff}$) were derived from the optical spectrum. The H$\delta$ and H$\gamma$ wings provided a measurement of $\log g$, while the He~{\sc i} 447-nm and He~{\sc ii} 454-nm transitions were used as the main diagnostic for $T_{\rm eff}$. However, fits to other helium transitions were also checked for consistency. Model fits to the He~{\sc ii} 454-nm and H$\gamma$ are shown in Fig.~\ref{parms_fig}. The uncertainties quoted in Table~\ref{params} already take into account the observed temporal variability (see Sect.~\ref{temp_var_sec}). The stellar luminosity $L_\star$ was assumed to be typical of other late O-type stars (see Martins et al. 2005a, Table 4). Therefore, through the analysis of IUE spectra we derived a distance of $\sim 1.3$~kpc and a corresponding reddening parameter $E(B-V)$ of 0.07 from the fit to the observed UV continuum. The obtained values are in good agreement with other modern literature estimates (e.g. Wegner 2002). The stellar radius was directly computed from $R_\star = (L_\star/4\pi \sigma T^4_{eff})^{1/2}$ and the stellar mass follows from $M_\star = gR_\star ^2/G$. Standard errors for $R_\star$ and $M_\star$ were computed assuming an error in the luminosity of $\pm$0.25 dex (see Martins et al. 2005b). Test models were also investigated with different abundances for the CNO elements, but precise values were difficult to determine and we found no clear evidence of chemical enrichment or depletion with respect to solar abundances of either Grevesse \& Sauval (1998) or the more recent Grevesse et al. (2007). A deeper study is necessary, but beyond the scope of this paper.

The wind parameters $\dot{M}$ and $v_\infty$ were determined from the fit to the C~{\sc iv} 154.8, 155.1-nm profile, from IUE data SWP21590, and appears consistent with all IUE data available. A standard $\beta$-velocity law was considered, with $\beta=1$. Due to the lack of wind diagnostic lines, clumping was not included in our analysis. Auger ionizations from X-rays were taken into account in the models, since they are essential in stars with spectral types similar to HD~57682 (see Martins et al. 2005b). The level of X-ray emission used in our CMFGEN models is taken from ROSAT measurements (Berghoefer et al. 1996), scaled to our distance.

\begin{table}
\centering
\caption{Summary of stellar and wind properties of HD~57682.}
\begin{tabular}{ll}
\hline
Spectral type             & O9IV (Walborn 1972)              \\
$T_{\rm eff}$ (K)             & 34 500 $\pm$ 1000 \\
log $g$ (cgs)             & 4.0 $\pm$ 0.2     \\
R$_{\star}$ (R$_\odot$)       & 7.0 $^{+2.4} _{-1.8}$ \\
$\log (L_\star/L_\odot)$    & 4.79 $\pm$ 0.25  \\
$M_{\star}$ ($M_{\odot}$)    & 17 $^{+19} _{-9}$ \\
$\log \dot{M}$ (M$_{\odot}$\,yr$^{-1}$) & -8.85 $\pm$ 0.50  \\
$v_{\infty}$ (km\,s$^{-1}$)   & 1200 $^{+500} _{-200}$  \\
$\log(L_X/L_{Bol})$ & -6.34 \\
\hline
\end{tabular}
\label{params}
\end{table}

\section{Temporal Variability and Rotation}\label{temp_var_sec}
\begin{figure}
\centering
\includegraphics[width=3.3in]{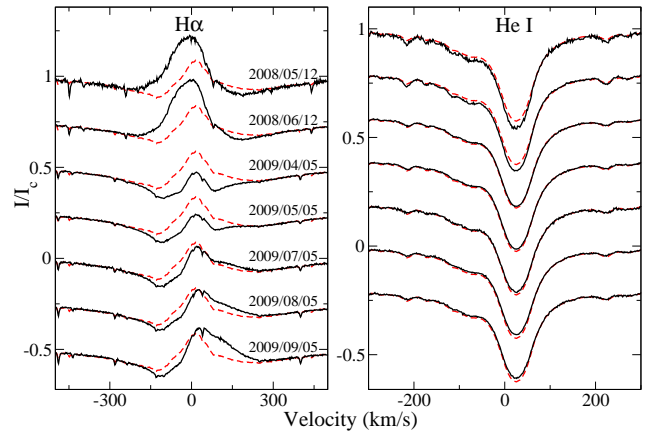}
\caption{Temporal variations of the H$\alpha$ (left) and He~{\sc i} 447-nm (right) lines throughout our observing run. The time-averaged profile is plotted in red (dashed) to emphasise variations. The date of observation is indicated in the H$\alpha$ frame.}
\label{temp_var}
\end{figure}

\begin{figure}
\centering
\includegraphics[width=3.3in]{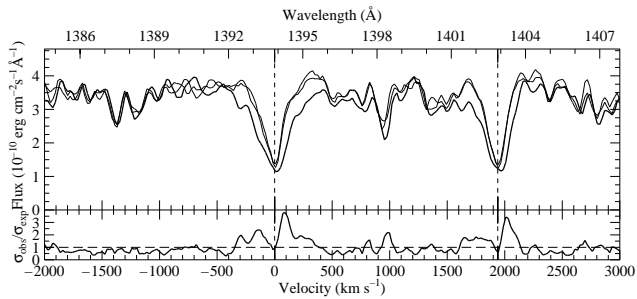}
\caption{Overplot of the Si~{\sc iv} line profiles of the three high-resolution IUE spectra of HD~57682. The two doublet rest wavelengths are indicated by vertical dashed lines. The lower part displays the significance of the variability, expressed as the square root of the ratio of the measured to the expected variances. Note that the variability observed at 1398~\AA\ is well-known and due to small differences in the echelle order overlap corrections.}
\label{iue_var}
\end{figure}

Our first observations obtained in 2008 Dec. show an enhanced H$\alpha$ emission and stronger absorption line depth in other elements compared to the observations obtained in 2009 May (see Fig.~\ref{temp_var} for an example with He lines). The stability of neighbouring lines and the use of a low-order polynomial during normalisation imply that the variability is real. An analogous anti-correlation is seen in the magnetic O-type star $\theta^1$~Ori~C (Donati et al. 2002; Wade et al. 2006). We also note that the longitudinal magnetic fields measured in 2008 Dec. versus 2009 May show opposite signs (see Table~\ref{obs_tab}). Additionally, the IUE UV spectroscopy shows significant variations of the 139.4, 140.2-nm Si~{\sc iv} doublet (see Fig.~\ref{iue_var}) - a phenomenon that is not observed in other UV lines and occurs exclusively in magnetic B-type stars, e.g. $\zeta$~Cas (Neiner et al. 2003) and all magnetic He-strong stars. Our CMFGEN models are not able to reproduce this variability by increasing the mass-loss rate, which, as we show in Sect.~\ref{disc_sec}, leads us to believe that it is due to magnetically-confined circumstellar material. 

In order to determine the rotational velocity $v\sin i$ of HD~57682 we computed the Fourier transform of many un-blended absorption lines throughout the optical spectrum (e.g. Gray 1981). The results of our analysis suggest that there is an important non-rotational contribution to the line profile. Our CMFGEN models employed an additional 40~km\,s$^{-1}$ macro-turbulence to provide a better fit to the observed absorption lines. From the analysis of about 30 different lines from the 2008 Dec. 6 and 2009 May 4 observations, we infer a mean $v\sin i$ of about $15\pm3$~km\,s$^{-1}$, in good agreement with Balona (1992). 

Due to the limited temporal sampling of our observations, we are unable to unambiguously determine the rotational period of HD~57682. However, using an upper limit on our inferred radius of 9.4~R$_{\sun}$ and the inferred rotational broadening of 15~km\,s$^{-1}$, we estimate that the maximum (rigid) rotation period $P_{\rm rot}$ is equal to about 31.5~d. The observed variations of H$\alpha$ (see Fig.~\ref{temp_var}) are qualitatively consistent with such a period, because the night-to-night changes are small compared to the difference between the mean profile corresponding to the two different observing runs. As well, the variability of the Stokes~$V$ profiles and longitudinal field measurements also support a variability timescale considerably longer than 5 nights. We therefore conclude that the variability of the magnetic and spectroscopic observables are consistent with rotational modulation, with a period of a few weeks.

\section{Magnetic Field}\label{mag_sec}
The longitudinal field measurements imply a strength of the dipole component of the magnetic field of at least 1~kG, consistent with the strengths of fossil fields studied in Ap/Bp stars, as well as other B and O-type stars. To estimate the surface magnetic field characteristics of HD~57682, we used the same method as Petit et al. (2008; and in prep.), which compares the observed Stokes~$V$ profiles to a large grid of synthetic profiles, described by an oblique-rotator model, to estimate the magnetic properties of the star. The model is parametrized by the dipole field strength $B_p$, the rotation axis inclination $i$, the positive magnetic axis obliquity $\beta$, and the rotation phase $\varphi$. In order to compute synthetic Stokes~$V$ profiles, we need to infer the line properties using the Stokes~$I$ profile. As the detailed formation of the line profile and the source of its variation is uncertain, we opted for the simplest assumption and used the mean of the observed profiles to determine the intensity profile parameters. 

\begin{figure*}
\centering
\includegraphics[width=6.5in]{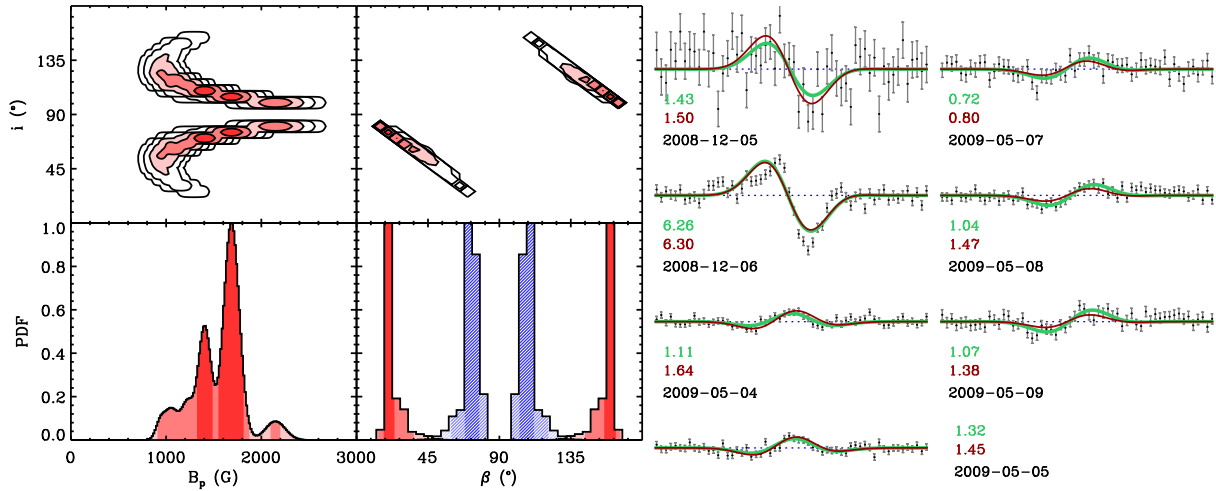}
\caption{{\bf Left:} Marginalised \textit{a posteriori} probability density for the dipole strength (bottom left), the magnetic field obliquity (bottom right and solid-filled red, with the inclination PDF overplotted in line-filled blue, the $B_{p}$-$i$ plane (top left) and the $i$-$\beta$ plane (top right). The 68.3\%, 95.4\% and 99.7\% credible regions are indicated in dark to light shades respectively. The additional contours for the 2D planes are 99.994\% and 99.9999\%. {\bf Right:} Our time series in Stokes~$V$ are shown, along with individual geometries that provide the maximum likelihoods for single observations (in thick green) and with the geometry that provides the maximum likelihood for all the observation combined (in thin red). The associated reduced~$\chi^2$ are also indicated.}
\label{sv_fits}
\end{figure*}

Assuming that only $\varphi$ may change between different observations of a given star, the goodness-of-fit of a given rotation-independent ($B_p$, $i$, $\beta$) magnetic configuration can be computed to determine a configuration that provides the overall best likelihood for all the observed Stokes~$V$ profiles. In Fig.~\ref{sv_fits}, we compare the synthetic profiles of the overall best likelihood configuration (in green) with the best likelihood configuration for each observation (in red). The quality of the different fits are similar, showing that a single inclined dipole can reproduce the observations as well as individual dipole configurations, although both are imperfect. However, any features that cannot be explained are treated formally as additional noise.

Fig.~\ref{sv_fits} shows the resulting marginalised {\it a posteriori} probability density function (PDF) for the dipole field strength (bottom left). The inferred value for the magnetic surface field strength in the 68.3\% credible region, is then $1680~^{+134}_{-356}$~G. The value of the dipole field strength will vary greatly with a small change of the inclination value, as illustrated in the $B_p$-$i$ 2D PDF shown in Fig.~\ref{sv_fits} (top left). A determination of the rotation period by further observations will improve the estimate of the dipole field strength.

Based on these data, the obliquity of the magnetic axis is closely correlated with the inclination as well, as demonstrated by the $\beta$-$i$ 2D PDF shown in Fig.~\ref{sv_fits} (top right). According to the 99.7\% credible regions of the marginalised PDFs of $i$ and $\beta$ (shown in Fig.~\ref{sv_fits} in blue and red respectively), we infer that the obliquity is between $10^\circ$ and $50^\circ$ if the inclination is between $47^\circ$ and $84^\circ$, or between $130^\circ$ and $169^\circ$ if the inclination is between $95^\circ$ and $132^\circ$.

\section{Discussion and Conclusions}\label{disc_sec}
It is interesting to note that the mass-loss rate derived from the H$\alpha$ profiles is inconsistent with mass-loss derived from the C~{\sc iv} UV lines. Test models indicate that the intensity of the H$\alpha$ emission observed at different dates requires mass-loss rates above $\sim$$10^{-7}$~M$_\odot$ yr$^{-1}$. With such high values, the predicted UV spectra would present strong P-Cygni profiles (e.g. Marcolino et al. 2009), which are not seen in any of the IUE spectra available for HD~57682. In fact, HD~57682 presents a UV spectrum typical of Galactic weak wind stars, with no significant P-Cygni emissions (Martins et al. 2005b; Marcolino et al. 2009). This leads us to believe that the H$\alpha$ profiles are not formed in the wind, but are possibly a result of confined circumstellar material. Walborn (1980) already reported this line to be unusual and of non-nebular origin.

A preliminary Bayesian analysis suggests that the magnetic field of HD~57682 is approximately reproduced by a dipole field with a characteristic surface field strength of $1680~^{+134}_{-356}$~G. We computed the magnetic wind confinement parameter (ud-Doula \& Owocki 2002) $1.4\times10^4$ for HD~57682, using $B_p\sim1600$~G, and the physical parameters listed in Table~\ref{params} ($\eta_*$ ranges from $4\times10^3$ - $2\times10^4$ using the 99.7\% credible regions). Although strongly confined, this places HD~57682 at an interesting intermediate regime amongst OB stars, between the super-strongly-confined wind of $\sigma$~Ori~E ($\eta_*\sim10^7$) and the much more weakly confined wind of $\theta^1$~Ori~C ($\eta_*\sim20$). Magnetic confinement of the wind of HD~57682 would naturally explain the H$\alpha$ variability, the UV variability, and the line depth variability. Moreover, shocks produced in a magnetically channelled wind could boost the X-ray emission, providing an explanation of the high X-ray luminosity without requiring a high mass-loss rate (e.g. Cohen et al. 2008). However, it is puzzling that HD~57682 presented a soft X-ray spectrum during the ROSAT observations, contrary to what would be expected from a fast wind and the strong magnetic confinement.

This discovery brings the number of firmly-established magnetic O-type stars to four. With its strong wind confinement, and its stellar and wind properties determined, HD 57682 provides a testbed for future MHD simulations. However, more data are needed to better determine the surface dipole field strength, magnetic obliquity, and inclination.

\section*{ACKNOWLEDGMENTS}
Some of the data presented in this paper were obtained from the Multimission Archive at the Space Telescope Science Institute (MAST). GAW and NSL acknowledge Discovery Grant support from the Natural Science and Engineering Research Council of Canada. JHG is supported by NSERC Discovery Grants held by GAW and David Hanes (Queen's University) and the Ontario Graduate Scholarship. WLFM and JCB acknowledge financial support from the French National Research Agency (ANR) through program number ANR-06-BLAN-0105. WLFM acknowledges the grant provided by IAU (Exchange of Astronomers Program) and CNES. DHC and RHDT are supported by NASA grant {\it LTSA}/NNG05GC36G.

\label{lastpage}
\end{document}